\input harvmac
 
\font\ensX=msbm10
\font\ensVII=msbm7
\font\ensV=msbm5
\newfam\math
\textfont\math=\ensX \scriptfont\math=\ensVII \scriptscriptfont\math=\ensV
\def\ensemble{\fam\math\ensX}

\font\srcX=rsfs10
\font\srcVII=rsfs7
\font\srcV=rsfs5
\newfam\scr
\textfont\scr=\srcX \scriptfont\scr=\srcVII \scriptscriptfont\scr=\srcV
\def\script{\fam\scr\srcX}

\font\rmptit=cmr5

\def\SV{{\script V}}
\def\SN{{\script N}}

\def\RR{{\ensemble R}}

\def\II{{\ensemble I}}

\def\ZZ{{\ensemble Z}}
\def\FF{{\ensemble F}}

\font\russeX=wncyr10


\def\CZ{{\cal Z}}
\def\CS{{\cal S}}
\def\CO{{\cal O}}
\def\CD{{\cal D}}

\def\CG{{\cal G}}
\def\CR{{\cal R}}
\def\CN{{\cal N}}

\def\CM{{\cal M}}




\def\da{{\dot \alpha}}
\def\Sl{S_\alpha}
\def\Sr{S_\da}
\def\chil#1{\chi_{#1}}
\def\chir#1{\bar\chi^{\dot#1}}
\def\huits{{\bf 8_s}}

\def\huitc{{\bf 8_c}}
\def\tr{{\rm tr}}
\def\Tr{{\rm Tr}}

\def\Definition{\mathrel{\hbox{\raise 5pt\hbox{\rmptit def}\kern -8pt =}}}
\def\d{{\rm d}}
\def\cyrf{\hbox{\russeX f}}


\parindent 25pt
\overfullrule=0pt
\tolerance=10000



\def\lr{\lref}

\def\npb#1(#2)#3 {Nucl. Phys. {\bf B#1} (#2) #3 }
\def\rep#1(#2)#3 {Phys. Rept.{\bf #1} (#2) #3 }
\def\plb#1(#2)#3{Phys. Lett. {\bf #1B} (#2) #3}
\def\prl#1(#2)#3{Phys. Rev. Lett.{\bf #1} (#2) #3}
\def\physrev#1(#2)#3{Phys. Rev. {\bf D#1} (#2) #3}
\def\ap#1(#2)#3{Ann. Phys. {\bf #1} (#2) #3}
\def\rmp#1(#2)#3{Rev. Mod. Phys. {\bf #1} (#2) #3}
\def\cmp#1(#2)#3{Comm. Math. Phys. {\bf #1} (#2) #3}
\def\mpl#1(#2)#3{Mod. Phys. Lett. {\bf #1} (#2) #3}
\def\ijmp#1(#2)#3{Int. J. Mod. Phys. {\bf A#1} (#2) #3}
\def\jhep#1(#2)#3{JHEP {\bf #1} (#2) #3}
\def\jmp#1(#2)#3{J. Math. Phys. {\bf #1} (#2) #3}
\def\adv#1(#2)#3{Adv. Theor. Math. Phys. {\bf #1} (#2) #3}

\lr\rfIntriligator{K. Intriligator, {\sl Bonus Symmetries of $N=4$ Super
    Yang-Mills Correlation Functions via AdS Duality}, hep-th/9811047.}
\lr\rfHalpern{M. Claudson and M.B. Halpern, {\sl Supersymmetric Ground State
    Wave Functions}, Nucl. Phys. {\bf B250} (1985) 689;\hfill\break
M. Halpern and C. Schwartz, {\sl Asymptotic Search for Ground States of SU(2)
    Matrix Theory}, Int. J. Mod. Phys. {\bf A13} (1998) 4367, hep-th/9712133.}
\lr\rfVanhoveWork{P.~Vanhove, {\sl Work in progress.}}
\lr\rfLercheElliptic{W.~Lerche, {\sl Elliptic Index and Superstring Effective
    Actions}, \npb308(1988)102.}
\lr\rfBanksSeibergSilverstein{T.~Banks, N.~Seiberg and E.~Silverstein, {\sl
    Zero and One-dimensional Probes with N=8 Supersymmetry}, \plb401(1997)30,
    hep-th/9703052.}
\lr\rfDanielssonFerretti{U.H.~Danielsson and G.~Ferretti, {\sl The Heterotic
    life of the D-particle}, \ijmp12(1997)4581, hep-th/9610082.}
\lr\rfBachasGreenSchwimmer{C.P.~Bachas, M.B.~Green and A.~Schwimmer, {\sl
    (8,0) Quantum mechanics and symmetry enhancement in type~I' superstrings},
    \jhep9801(1998)006, hep-th/8712086.}
\lr\rfGutperleIprime{M.~Gutperle, {\sl A note on heterotic/type I' duality and
    D0-branes quantum mechanics}, hep-th/9903010.}
\lr\rfDhokerPhongFermions{E.~D'Hoker and D.H.~Phong, {\sl Loop Amplitudes for
    the Fermionic String}, \npb278(1986)225.}
\lr\rfThompson{G.~Thompson, {\sl 1992 Trieste Lectures on Topological Gauge
    Theory and Yang-Mills Theory}, hep-th/9305120.}
\lr\rfKiritsisObers{E.~Kiritsis, N.A.~Obers, {\sl Heterotic/Type~I duality in
    $D<10$ dimensions}, \jhep9710(1997)004, hep-th/9709058.}
\lr\rfBFKOV{C.~Bachas, C.~Fabre, E.~Kiritsis, N.A.~Obers and P.~Vanhove, {\sl
    Heterotic/Type~I duality and D-brane instantons}, \npb509(1988)33,
    hep-th/9707126.}
\lr\rfSwanseaAdS{N.~Dorey, T.J.~Hollowood, V.V.~Khoze, M.P.~Mattis and
  S.~Vandoren, {\sl Multi-instantons and Maldacena's Conjecture},
  hep-th/9810243, {\sl Multi-Instanton Calculus and the AdS/CFT Correspondence
    in N=4 Superconformal Field Theory}, hep-th/9901128.} 
\lr\rfAdSCFT{J.~Maldacena, \adv2(1998)231, hep-th/9711200; E.~Witten,
  \adv2(1998)253, hep-th/9802150; I.~Klebanov and A.A.~Tseytlin,
  \npb475(1996)179, hep-th/9604166; S.~Gubser and I.~Klebanov,
  \plb413(1997)41, hep-th/9708005; S.~Gubser, I.~Klebanov and A.~Polyakov,
  \plb428(1998)105, hep-th/9802109.} 
\lr\rfBillo{M.~Bill{\'o}, M.~Caselle, A.~D'Adda and P.~Provero, {\sl Matrix string
    states in pure 2d Yang Mills theories}, hep-th/9809095; {\sl 2D
    Yang-Mills Theory as a Matrix String Theory}, Talk presented at the 2nd
  Conference on Quantum aspects of Gauge Theories, Supersymmetry and
  Unification, Corfu, Greece, 21-26 September 1996, hep-th/9901053.} 
\lr\rfVanhovePhd{P.~Vanhove, {\sl Au bout de la corde...la th{\'e}orie M},
  PhD thesis, {\'E}cole Polytechnique April 1998.}
\lr\rfDVVproceeding{R.~Dijkgraaf, E.~Verlinde and H.~Verlinde, {\sl Notes on
    Matrix and Micro Strings}, Nucl. Phys. Proc. Suppl. {\bf 68} (1998) 28,
    hep-th/9709107.} 
\lr\rfFujikawa{K.~Fujikawa, {\sl Path-integral Measure for Gauge-invariant
  Fermion Theories}, \prl42(1979)1195; {\sl Path integral for gauge theories
  with fermions}, \physrev21(1980)2848;
  \physrev22(1980)1499 (Erratum); \physrev23(1981)2262.}
\lr\rfGiddingsVerlindeHacquebord{S.B.~Giddings, F.~Hacquebord, H.~Verlinde,
    {\sl High Energy Scattering and D-Pair Creation in Matrix String Theory},
    \npb537(1999)260, hep-th/9804121.} 
\lr\rfSethiSternPaban{S.~Paban, S.~Sethi, M.~Stern, {\sl Summing Up
    Instantons in Three-Dimensional Yang-Mills Theories}, hep-th/9808119.} 
\lr\rfSeibergOrbifold{N.~Seiberg, {\sl Note on Theories with 16 supercharges},
  Nucl.Phys.Proc.Suppl. 67 (1998) 158, hep-th/9705117.}
\lr\rfHorava{P.~Ho{\v r}ava, {\sl Matrix Theory and Heterotic Strings on
    Tori}, \npb505(1997)84, hep-th/9705055.} 
\lr\rfRey{S.J.~Rey, {\sl Heterotic M(atrix) Strings and Their Interactions},
    \npb502(1997)170, hep-th/9704158.} 
\lr\rfLowe{D.A.~Lowe, {\sl Heterotic Matrix String Theory}, \plb403(1997)243,
    hep-th/9704041.} 
\lr\rfLang{S.~Lang, {\sl Algebra}, page 53, third edition, Addison-Wesley
    Publishing Company, 1993.} 
\lr\rfPolchinskiDbrane{J.~Polchinski, {\sl Dirichlet-Branes and Ramond-Ramond
    Charges}, \prl75(1995)4724, hep-th/9510017.}
\lr\rfSimon{B.~Simon, {\sl Some quantum Operators with Discrete Spectrum but
  Classically Continuous Spectrum}, \ap146(1983)209.}
\lr\rfBorel{A.~Borel, {\sl Automorphic forms on $Sl(2,\RR)$}, Cambridge Tracts
  in Mathematics 1997.}
\lr\rfHoweWest{P.S.~Howe and P.C.~West, {\sl The complete N=2 D=10
  supergravity}, \npb238(1984)181.}
\lr\rfSchwarzIIb{J.H.~Schwarz, {\sl Covariant field equations of chiral N=2
    D=10 supergravity}, \npb226(1983)269.}
\lr\rfdeWitHoppeNicolai{B.~de Wit, J.~Hoppe, H.~Nicolai, {\sl  On the Quantum Mechanics of Supermembranes}, \npb305(1988)545}
\lr\rfdeWitLouis{B.~de Wit and J.~Louis, {\sl Supersymmetry and Dualities in
    various dimensions}, hep-th/9801132.}
\lr\rfPlefkaWaldron{J.~Plefka and A.~Waldron, {\sl On the Quantum Mechanics of
    M(atrix) Theory}, \npb512(1998)460,
    hep-th/9710104; {\sl Asymptotic Supergraviton States in Matrix Theory},
    hep-th/9801093.} 
\lr\rfdeWitLuscherNicolai{B.~de Wit, M.~L{\"u}\unskip scher and H.~Nicolai, {\sl
    The Supermembrane is Unstable},  \npb320(1989)135.} 
\lr\rfKostovVanhove{I.K.~Kostov and P.~Vanhove, {\sl Matrix String Partition
  Functions}, \plb444(1998)196, hep-th/9809130.}
\lr\rfMooreNekrasovShatashvili{G.~Moore, N.A.~Nekrasov, S.~Shatashvili, {\sl
  D-particle bound states and generalised instantons}, hep-th/9803265.}
\lr\rfIKKT{N.~Ishibashi, H.~Kawai, Y.~Kitazawa and A.~Tsuchiya, {\sl A Large-N
    Reduced Model as Superstring}, \npb498(1997)467, hep-th/9612115.} 
\lr\rfKrauthNicolaiStaudacher{V.~Krauth, H.~Nicolai and  M.~Staudacher, {\sl
    Monte Carlo Approach to M-theory}, \plb431(1998)31, hep-th/9803117;
    V.~Krauth and M.~Staudacher, {\sl Finite Yang-Mills Integrals},
    \plb435(1998)350, hep-th/9804199; {\sl Eigenvalue Distributions in
    Yang-Mills Integrals}, hep-th/9902113}  
\lr\rfGreenRomeII{M.~Bianchi, M.B.~Green, S.~Kovacs and G.~Rossi, {\sl
    Instantons in supersymmetric Yang-Mills and D-instantons in IIB
    superstring theory}, \jhep9808(1998)13, hep-th/9807033}
\lr\rfGreenGutperleKwon{M.B.~Green, M.~Gutperle and H-K.~Kwon, {\sl
    Sixteen-Fermion and Related terms in M-theory on $T^2$}, \plb421(1998)149,
  hep-th/9710151.} 
\lr\rfGibbonsGreenPerry{G.W.~Gibbons, M.B.~Green and M.V.~Perry, {\sl
    Instantons and seven-branes in type IIB superstring theory},
    \plb370(1996)37, hep-th/9511080.}  
\lr\rfWittenBound{E.~Witten, {\sl Bound States Of Strings And $p$-Branes},
    \npb460(1996)335, hep-th/9510135}  
\lr\rfSethiGreen{S.~Sethi and M.B.~Green, {\sl Supersymmetry Constraints on
    Type IIB Supergravity}, hep-th/9808061.} 
\lr\rfTaylorCompact{W.~Taylor, {\sl D-brane field theory on compact spaces},
  \plb394(1997)282, hep-th/9611042.} 
\lr\rfTaylorLecture{W.~Taylor, lectures presented at Trieste
   summer school on particle physics and cosmology, June 1997, hep-th/9801182.}
\lr\rfBonora{G.~Bonelli, L.~Bonora, F.~Nesti, {\sl String Interactions from
    Matrix String Theory}, \npb538(1999)100, hep-th/9807232.}
\lr\rfWynter{T.~Wynter, {\sl Anomalies and large N limits in matrix string
    theory}, \plb415(1997)349, hep-th/9709029.}
\lr\rfGreenGutperleDinstantons{M.B.~Green and M.~Gutperle, {\sl Effects of
    D-instantons}, \npb498(1997)195, hep-th/9701093.}
\lr\rfYi{P.~Yi, {\sl Witten Index and Threshold Bound States},
  \npb505(1997)307, hep-th/9704098} 
\lr\rfSethiStern{S.~Sethi and M.~Stern, {\sl D-brane Bound State Redux},
  \cmp194(1998)675, hep-th/9705046.} 
\lr\rfWittenIndex{E.~Witten, {\sl Constraints on Supersymmetry Breaking},
  \npb202(1982)253.} 
\lr\rfBFSS{T.~Banks, W.~Fischler, S.~Shenker, and L.~Susskind,
   {\sl M Theory As A Matrix Model: A Conjecture}, \physrev55(1997)5112,
   hep-th/9610043.} 
\lr\rfThompson{G.~Thompson, {\sl Topological Gauge Theory and Yang-Mills
    Theory\/}, Trieste School 1992, hep-th/9305120.}
\lr\rfBaake{M.~Baake, P.~Reincke and V.~Rittenberg, {\sl Fierz identities for
    real Clifford algebras and the number of supercharges}, \jmp26(1985)1070}
\lr\rfDVV{R.~Dijkgraaf, E.~Verlinde  and H.~Verlinde, {\sl Matrix String
    Theory}, \npb500(1997)43, hep-th/9703030.}
\lr\rfGreenGutperleTwo{M.B.~Green and M.~Gutperle, {\sl Configurations of two
    D-instantons}, \plb398(1997)69, hep-th/9612127.} 
 
\rightline{hep-th/9903050}
\rightline{DAMTP-1998-156}
\Title{}
{\vbox{\centerline {D-instantons and Matrix Models}
}}
 
\centerline{Pierre Vanhove}
\centerline{{\tt p.vanhove@damtp.cam.ac.uk}}

\centerline{\it Department of Applied Mathematics and Theoretical Physics}
\centerline{\it  Cambridge University}
\centerline{\it Cambridge CB3 9EW, UK}

\vskip .3in

{
We discuss the Matrix Model aspect of configurations saturating a fixed number
of fermionic zero modes. This number is independent of the rank of the gauge
group and the instanton number. This will allow us to define a large-$N_c$
limit of the embedding of $K$ D-instantons in the Matrix Model and make
contact with the leading term (the measure factor) of the supergravity
computations of D-instanton effects. We show that the connection between
these two approaches is done through the Abelian modes of the Matrix
variables. 
}
 
\Date{March 1999}



\newsec{Introduction}

Over the past four years some tremendous progress and insights about the
non-perturbative and global behaviour of supersymmetric gauge theory,
superstring theory and supergravity have appeared. All these advances are
founded on a web of consistent cross-checked conjectures culminating with the
idea of M-theory as the mother of all theories. Most of the impressive and
exact non-perturbative results were derived by considering BPS saturated
amplitudes. Due to the saturation of the fermionic zero-modes these terms are
protected by some non renormalisation theorems and can be computed both in the
perturbative and non-perturbative regimes. This is the case of the
eight-fermion terms in the three-dimensional super Yang-Mills theory
\refs{\rfSethiSternPaban}, the D-instantons of the type~IIb string theory
\refs{\rfGreenGutperleDinstantons}, or the wrapped D1-brane around tori of
dimensions smaller than five in the type~I theory \refs{\rfBFKOV}. In these
cases, the D-instantons contributions belong to a half-BPS multiplet of the
theory, and they come from amplitudes where sixteen or eight fermionic
zero-modes, respectively for the type~IIb and type~I theory, have to be soaked
up. The surprising aspect of these results is that even in a vacuum containing
$K$ D-instantons it is only necessary to saturate the same fixed number of
fermionic zero modes, independently of $K$. This is because of the existence
of threshold bound states of D0-branes and the action of T-duality which
exchange $K$ D$p$-branes on top of each other singly wrapped around a
$(p+1)$-torus with one D$(p-1)$-brane wrapped $K$-times around a $p$-torus, and
the fact that the presence of winding modes does not break
supersymmetry.\foot{It should be noted that is not true for bound states of
  D-particle and anti-D-particle.}  

In the context of the correspondence between the supergravity results and the
CFT computation, this independence of fermionic zero-modes with respect to the
instanton number becomes much more obscure. Let us consider for example the
case of a vacuum containing $K$ D3-branes: it was claimed in \refs{\rfAdSCFT}
and impressively strengthen by the result of \refs{\rfSwanseaAdS} that the large-$N_c$ limit of
this theory is in correspondence with the four-dimensional super Yang-Mills
theory with gauge group $SU(N_c)$ in the large-$N_c$ 't Hooft limit ($N_c
g^2_{YM}=constant$). This confirmation used a sector of the theory with
a fixed number of fermionic zero modes independently of $N_c$.
The puzzle is that in the super Yang-Mills case one would normally
think that when the rank of the gauge group increases there are extra
fermionic zero-modes and the result cannot match the supergravity ones. In
fact it was understood by \refs{\rfSwanseaAdS} that an embedding of a
configuration of $K$ instantons in the $SU(N_c)$ group has a fixed number of
fermionic zero-modes independent of the number of instantons and the rank of
the group. 

The main purpose of the present article, is to explain that in the context of
Matrix models, such a configuration of fermionic zero-modes can be 
realized, and can lead to a way of defining a large-$N_c$ limit of the Matrix
model.  

In section~2, we will introduce the supergravity aspect of the D-instanton
effects, and in section~3 we will introduce the Matrix model that we will use
in the following. Section~4 contains a discussion of the dynamics of these
models, emphasing the importance of considering a gauge-invariant model. In
section~5 we compute the partition function of the Matrix model with various
symmetries, and map these results to the supergravity results in
section~6. Section~7 contains a discussion of this approach. Appendix~A
contains an explicit computation of the partition function of the
supersymmetric Matrix model with two real supercharges and Appendix~B
summarises our conventions for the $\Gamma$ matrices used in the text.

\newsec{The Supergravity Side}
\seclab\secsugra

The type~IIb chiral version of the ten dimensional supergravity is peculiar in
several respects. First, being chiral with maximum number of supersymmetries,
with two sets of sixteen real component supercharges of the same space-time
chirality, defined with respect to the projector $(1\pm \Gamma^{11})/2$, in ten
dimensions it has a richer moduli space than its non chiral counter part the
type~IIa supergravity. The superspace formalism of this chiral supergravity
theory was worked out in \refs{\rfHoweWest} and will be used here. The
superspace formalism uses a supermanifold with ten even coordinates, $x^\mu $
$(\mu =0,\cdots,9)$, and sixteen odd complex coordinates,\foot{These
  coordinates will be put in correspondence with the zero modes of the
  matricial fermions (see section~6.2).} $\theta^\alpha$
$(\alpha=1,\cdots,16)$, and their complex conjugates
$(\theta^\alpha)^*\Definition \theta^{\bar\alpha}$, the whole set packaged in
$z^M\Definition (x^\mu ,\theta^\alpha,\theta^{\bar\alpha})$. At each point in
the superspace there are some local coordinates related to the one-form $\d
z^M$ by the vielbein

$$
\d E^A \Definition \d z^M E_M^A,
$$
and as for usual Riemann manifolds the vielbein are invertible,

$$
E_A^M \, E_M^B = \delta_A^B.
$$
The tangent space is described by the covering of the group $SO(1,9)\times 
U(1)_B$. The $U(1)_B$ factor is a local phase transformation on the fermionic
coordinates by 

$$
\theta^\alpha \to \exp\left({i\over 2}\Gamma^{11}\xi\right)\theta^\alpha,\quad
\theta^{\bar\alpha} \to \exp\left(-{i\over
    2}\Gamma^{11}\xi\right)\theta^{\bar\alpha}. 
$$
It was shown in \refs{\rfHoweWest} that this $U(1)_B$ factor is precisely the
factor appearing in  the coset space parametrisation $SU(1,1)/U(1)_B \equiv
Sl(2,\RR)/U(1)_B$ of this theory. The $Sl(2,\RR)$ is a rigid transformation
acting on the left of the fields and the $U(1)_B$ induced by the chiral nature
of the theory is a local transformation acting on the right of the
coset. 
As in \refs{\rfSchwarzIIb} we parametrise this coset with the matrix
$$
V\Definition {1\over\sqrt{-2i\rho_2}} \pmatrix{
\rho e^{i\Xi}& \bar\rho e^{-i\Xi}\cr
e^{i\Xi}&e^{-i\Xi}\cr
}.
$$
The $Sl(2,\RR)$ group acts by a matrix multiplication on the left of this
matrix and the scalar $\rho$ transforms by a fractional linear transformation
$\rho\to (a\rho+b)/(c\rho+d)$. The $U(1)_B$ acts on the right by 

$$
V \to  V \pmatrix{
e^{i\xi}&0\cr
0&e^{-i\xi}\cr
},
$$
with
$$
e^{i2\xi}={ c\bar\rho+d\over c\rho+d}
$$

Using the supergravity equations of motion allows us to relate the field
$\rho$ to the Ramond scalar field $C^{(0)}$ and the dilaton and the string
coupling constant $g_s=\exp(\phi)$ as

$$
\rho=C^{(0)}+{i\over g_s}.
$$

The second special feature of this chiral theory resides in its peculiar
point-like solitonic solution to the equations of supergravity. Despite the
fact that this
solution looks singular, because it is localised in space-time, it belongs to
the class of D-brane instantonic solutions \refs{\rfPolchinskiDbrane}.
The metric induced by the presence of $N$ D-instantons is given (in Euclidean
space) in the Einstein metric \refs{\rfGibbonsGreenPerry} by
$$
g_{\mu \nu}^E = \delta_{\mu \nu},\quad
e^\phi = g_s \left(1+c_o {g_sNl_s^8\over r^8} \right).
$$

The presence of this D-instanton induces a correction of order
$\alpha'{}^3=l_s^6$ to the effective action expressed in the Einstein frame as
\refs{\rfGreenGutperleDinstantons}
\eqn\eAction{
\CS\sim {1\over l_s^8} \int d^{10}x \det(E^m_\mu ) \left[ R +
{l_s^6\over 2^{11} \pi^7} \left(
 f^{(0,0)}(\rho,\bar\rho) 
 {\cal R}^4 + f^{(12,-12)}(\rho,\bar\rho) \Lambda^{16}  + \cdots
\right) \right],
}
where  $\Lambda$ is a complex chiral
$SO(9,1)$ spinor which transforms under the $U_B(1)$ R-symmetry with charge
3/2 (see \refs{\rfGreenGutperleKwon} for detailed expressions).  
Quantum effects induced by looping around an arbitrary number of D-instantons are
given by the functions $f^{(w,-w)}(\rho,\bar\rho)$. 
These functions are modular forms of $Sl(2,\ZZ)$ of
 indicated weight up to a phase \refs{\rfBorel} $(\gamma\in Sl(2,\ZZ))$
$$
f^{(w,-w)}(\gamma\cdot \rho,\gamma\cdot \bar\rho)= \left( c\bar\rho+d\over
  c\rho+d\right)^w \ f^{(w,-w)}(\rho,\bar\rho).
$$
They are connected to the modular function
 $f^{(0,0)}$ by repeated action of the covariant derivative  ${\rm
   D}=(i\rho_2\partial_\rho + w/2)$ which 
maps a $(q,p)$ modular form into a $(q+1,p-1)$ form
\refs{\rfGreenGutperleKwon,\rfSethiGreen}. 
The small coupling expansion, $g_s\to 0$, of $f^{(w,-w)}$ reads 
$$
f^{(w,-w)} = 2\zeta(3)e^{-3\phi/2}- {\Gamma(-1/2+w)\over\Gamma(3/2+w)}\zeta(2)
e^{\phi/2} + \sum_{K=1}^\infty \CG_{K,w} e^{\phi/2},   
$$
where the first two terms have the form of string tree-level and one-loop
terms and $\CG_{K,w}$ contains the charge-$K$ D-instanton and anti-D-instanton
terms. The instanton contribution to $\CG_{K,w}$ has  the asymptotic expansion
in powers of $Ke^{-\phi}$,
\eqn\eModularexp{
\CG_{k,w} = \mu (K)(4\pi K e^{-\phi})^{-7/2} \left(e^{2i\pi K \rho} S^D_w + e^{-2i\pi K \bar\rho} S^D_{-w}
\right) \ .} 
We will denote hereafter 
\eqn\eMu{
\mu (K)=\sum_{0<m|N} {1\over m^2}
}
the measure factor, and the coefficients $S^D_w$ for the D-instantons and
$S^D_{-w}$ for the anti-D-instantons are given by 

\eqn\eCoeff{
\eqalign{
S^D_w =&(4\pi K e^{-\phi})^{w+4}\times  (-\sqrt{2}\Gamma(-1/2-w))\cr
&\left( 1+
  \sum_{p\geq 1} {(-1)^p\over p!}\times  
{\Gamma(p-w-1/2)\over\Gamma(-w-1/2)}\times {\Gamma(p-w+3/2)\over
  \Gamma(-w+3/2)}\times (4\pi K e^{-\phi})^{-p} \right)
}}

The phase of these modular functions is compensated by those of the fields
multiplying them in the action. These fields carry a tensorial structure which
has a well defined $U(1)_B$ weight  induced by the coset space structure of
the theory \refs{\rfSethiGreen}. As was discovered in
\refs{\rfHoweWest,\rfSchwarzIIb}, all these terms can be packaged in a chiral
superfield $\Phi(z^M)$ for the type~IIb supergravity, satisfying the
constraints,  
$$
D\Phi=0, \quad D^4\Phi=0=\bar D^4\Phi.
$$
These constraints imply that this superfield is independent of
$(\theta^\alpha)^*$ and is a function $\Phi(x^\mu -\bar\theta \gamma^\mu 
\theta,\theta^\alpha)$, which can be expanded in terms of $\theta$ and
$\bar\theta^*=\theta^T \Gamma^0$ as \refs{\rfHoweWest,\rfSchwarzIIb}

\eqn\ePhi{
\Phi = \tau_o+\hat\Phi=\tau - 2i\bar \theta^* \lambda -{1\over 24} \hat G_{\mu \nu\rho}
\bar\theta^* \gamma^{\mu \nu\rho} \theta + \cdots -{i\over 48}  \CR_{\mu \sigma\nu\tau}
  \bar\theta^* \gamma^{\mu \nu\rho} \theta \bar\theta^*
\gamma^{\sigma\tau}_{\ \ \ \rho} \theta +\cdots
}
and the correction to the action can be obtained by picking the terms containing
sixteen powers of $\theta$ in the Taylor expansion of some unknown function
of the chiral superfield $\Phi$~\refs{\rfSethiGreen}
$$
\CS^{(3)} = l_s^{-2} \int d^{10}x d^{16}\theta \det(E_\mu ^m)\
\left(F[\Phi]+c.c. \right).
$$

The superfield $\Phi$ does not have a well defined $U(1)_B$ weight,
because $\tau$ does not have a proper weight
\refs{\rfIntriligator}. But the fluctuations $\delta\tau$ around a
classical value $\tau_o$ have weight $+2$ like the other terms
in $\hat\Phi$.\foot{$P_\mu =\partial_\mu \Phi$ with
  $P_\mu =-\varepsilon_{\alpha\beta}V_+^\alpha\partial_\mu V^\beta_+$ has
  $U(1)_B$ weight $+2$. The weights normalized as in \refs{\rfSchwarzIIb} are half the ones of \refs{\rfHoweWest,\rfIntriligator}}
The superfield only depends on half of the fermionic coordinates. This is
natural as it is related to the contribution of the D-instantons, which are
half-BPS states. This welcome feature will enable us to find a correspondence
to this field in the Matrix Model in section~6.2.
We will now turn to an interpretation of the previous formulae in the Matrix
model setting.

\newsec{The Matrix Model}

The D$p$-brane, and in particular the D-instantons $(p=-1)$, have their
dynamics described by open strings attached on their world-sheet
\refs{\rfPolchinskiDbrane}; thus the low energy excitations of $N$ D$p$-branes
on top of each other are described by a dimensional reduction to $p+1$
dimension of the ten dimensional super Yang-Mills theory with gauge group
$U(N)$ \refs{\rfWittenBound} (see \refs{\rfTaylorLecture} for a comprehensive
lecture on this subject). The ten dimensional super Yang-Mills theory has the
Lagrangian 
$$
\CS_{[D=10]} = {1\over g^2_{10}} \, \int d^{10}x  \Tr\left(-{1\over 4} F^2 +{i\over 2}
\Psi^T\Gamma^0\Gamma^\mu D_\mu \Psi\right),
$$
where the curvature field $F_{\mu \nu}= [D_\mu ,D_\nu]$, the covariant
derivative $D_\mu \Definition \partial_\mu  -i [A_\mu , \cdot ]$, the Hermitian
connection $A_\mu = A_\mu ^a T_a$ and the fermion fields are sixteen component
Majorana-Weyl spinors of $SO(1,9)$  Hermitian matrices in the adjoint of the
gauge group $\Psi = \Psi^a T_a$. This theory is invariant under the
supersymmetry transformation 

$$
\delta_\epsilon A_\mu ^a = {i\over2} \bar\epsilon \Gamma_\mu \Psi^a,\quad
\delta_\epsilon \Psi^a  = - {1\over 4} \Gamma^{\mu \nu} F^a_{\mu \nu}
\epsilon.
$$
$\epsilon$ is a Majorana-Weyl spinor.

For later convenience, we rewrite the fields and the supersymmetry
transformation by splitting the $SU(N)$ and the Abelian $U(1)$ part of the
matrices: $\Psi \Definition \psi + \theta \II$, $A_\mu \Definition X_\mu +
x_\mu \II$, $F_{\mu \nu}\Definition F_{\mu \nu}+ f_{\mu \nu} \II$ and
$$
\delta_\epsilon^1 X_\mu ^a = {i\over2} \bar\epsilon \Gamma_\mu \psi^a,\quad
\delta_\epsilon^1 \psi^a  = - {1\over 4} \Gamma^{\mu \nu} F^a_{\mu \nu}
\epsilon,
$$
for the $SU(N)$ part and
$$
\delta_\epsilon^2 x_\mu = {i\over2} \bar\epsilon \Gamma_\mu \theta,\quad
\delta_\epsilon^2 \theta = - {1\over 4} \Gamma^{\mu \nu} f_{\mu \nu}
\epsilon,
$$
for the $U(1)$ part.

In the particular case of the D-instanton we get a zero-dimensional model,
({\sl i.e.}, the variables no longer depend on any coordinates)
called the IKKT Matrix Model \refs{\rfIKKT,\rfWittenBound}.
The Lagrangian reduces to
\eqn\eDefIKKT{
\CS_{[10\to 0]} = {1\over g^2_{0}} \Tr \left({1\over 4} \sum_{0\leq \mu <\nu\leq
    9} [X_\mu ,X_\nu]^2+{1\over 2} \psi^T \Gamma^0\Gamma^\mu [X_\mu ,\psi]\right), 
}
$$
\delta_\epsilon^1 X_\mu ^a = {i\over2} \bar\epsilon \Gamma_\mu \psi^a,\quad
\delta_\epsilon^1 \psi^a  = - {1\over 4} \Gamma^{\mu \nu} F^a_{\mu \nu}
\epsilon,
$$
where $\epsilon$ is a Majorana-Weyl spinor.
The model possesses an additional supersymmetry transformation which only
involves the $U(1)$ part of the fields
\eqn\eAbeliansuzy{
\delta^2_\zeta \theta = \zeta, \quad \delta^2_\zeta x_\mu = {i\over 2}
\bar\zeta \Gamma^\mu \theta,
}
with $\zeta$ a Majorana-Weyl spinor. As all the fields are in the adjoint of
the group, the Abelian piece of the coordinates has disappeared in the previous
action, but as we will see in section~6.2 this part still plays a role in the
dynamics of the model. It should be noted that this supersymmetry
transformation is a superspace translation for the coordinate $z^M$.

As this model gives a description of the low-energy excitation of the open
string with end points fixed on the D-instantons, the gauge coupling of this
model is fixed to be 
$ g^2_{0} =g_s/l_s^4. $

Two other models relevant to this article are: (1) the dimensional reduction to
the Quantum mechanical model in $1+0$ dimensions
\refs{\rfHalpern,\rfdeWitHoppeNicolai,\rfdeWitLuscherNicolai}, known as the
BFSS model after its revival by the article \refs{\rfBFSS} 
\eqn\eDefBFSS{
\eqalign{
\CS_{[10\to 1]} = {1\over 2g^2_1} \, \int dt&  \Tr\left(-{1\over 2} (\partial_t
  A^m)^2 +{i\over 2\pi l_s^2} \Psi^T \Gamma^0\partial_t \Psi+\right.\cr
 &\left.{1\over 4\pi^2 l_s^4} \sum_{1\leq m<n\leq 9} [X_m,X_n]^2 + {1\over 4\pi^2 l_s^4} \psi^T\Gamma^0\Gamma^m[X_m\psi]\right),
}}
where the coupling constant is given by $g^2_1=g_sl_s$;
and (2) the dimensional reduction to $1+1$ dimensions considered by \refs{\rfDVV}, 
\eqn\eDefDVV{
\eqalign{
\CS_{[10\to 2]} = {1\over g^2_2} \, \int d^2x & \Tr\left(-{1\over 4} F^2 -
  {1\over 8\pi^2 l_s^4}(D_aA_I)^2+ {i\over 4\pi l_s^2} \Psi^T\Gamma^0\Gamma^aD_a\Psi+\right.\cr
  &\left.{1\over 8\pi^2 l_s^4} \sum_{1\leq I<J\leq 8} [X_I,X_J]^2  +{1\over
      8\pi^2 l_s^4} \psi^T \Gamma^0\Gamma^I[X_I,\psi]\right),\cr
}}
the coupling constant is given by $g^2_2=g_s(2\pi l_s)^{-2}$.

In order to stress the importance of the supersymmetry we will consider more
generally the models deduced by dimensional reduction from the $D=3,4,6$ and 10
super Yang-Mills theory. The amplitudes associated with these half-BPS
contributions correspond to the vacuum expectation values of the same number
of fermionic zero modes as the non-broken supersymmetries. That is, sixteen
real fermions for $D=10$, eight real fermions for $D=6$ and four real fermions
for $D=4$. This product of fermionic zero modes is the fermion number operator
of the theory $(-)^F$. So this amplitude is (a part of) the Witten index of
the model.

\newsec{The Dynamics of the Models}

These matrix models all have in common a quartic bosonic potential $V_B = \Tr
[X,X]^2$ and a fermionic one $V_F = \Tr \Psi^T \Gamma^0\Gamma^\mu [X_\mu 
,\Psi]$. As the coordinates are Hermitian matrices, the potential $V_B$ is
negative definite. The classical space of configurations is given by the
vanishing of this potential $V_B=0=V_F$, and is described by the space
parametrised by the eigenvalues of the matrices modulo permutations, so this is

$$
\CM=\left(\RR^d \right)^N/S_N,
$$
if we have $d$ matrix coordinates $X^i$ $(i=1,\dots,d)$ in the adjoint of
$U(N)$ whose Weyl group is $S_N$. The potential is composed of valleys along
the direction in the Cartan subalgebra of the group with a harmonic oscillator
like shape $V_B\sim \omega^2 y^2$ for the coordinates orthogonal to these
directions. 

The description of this system in terms of harmonic oscillators shows that the
spectrum of the {\it purely bosonic theory\/} is discrete (see \refs{\rfSimon}
for a mathematical proof, \refs{\rfKrauthNicolaiStaudacher} for numerical
evidence). The physical argument given in \refs{\rfdeWitLuscherNicolai}, is
that one chooses a flat direction by setting the coordinate $X^d$ in the Cartan
subalgebra of the group; then all transverse coordinates have a harmonic
potential $V_B\sim |X^d|^2 Y^2$. The quantum model has a zero-point energy
$E_0 \sim |X^d|/2$ which grows linearly along the flat direction,
preventing the wave function from extending to infinity along $X^d$ and
localising it. It then follows that the spectrum is discrete. As was
rigorously shown by de Wit, L{\"u}scher and Nicolai in
\refs{\rfdeWitLuscherNicolai}, the spectrum of the supersymmetric theory is
continuous. This is due to the cancellation of the zero energy between the
transverse fluctuations of the bosonic coordinates orthogonal to the Cartan
directions and their fermionic partners. Their result showing that there is
continuous spectrum, does not prevent the theory from having a discrete
spectrum sitting in the middle of it. A non-zero value for the Witten Index of
the theory will show that. The computation of the index is complicated by the
presence of the continuous spectrum and the flat directions which could lead to
infrared divergences. But as noticed from explicit computation
\refs{\rfGreenGutperleTwo,\rfYi,\rfSethiStern}, this is not the case. We recall
briefly how to perform the computation of the Witten index in order to justify
the infrared finiteness of the model we will consider later on.

The Witten index of the the $U(N)$ supersymmetric quantum mechanical model in
$D-1$ dimensions (\eDefBFSS\ is the particular case $D=10$) is defined as the
trace with insertion of the fermion counting operator

$$I_W\Definition \lim_{\beta\to\infty} \tr\left( (-)^F e^{-\beta H}\right).$$

It was shown by Yi \refs{\rfYi} and Sethi and Stern \refs{\rfSethiStern}, that
the
computation can be reduced to computing the bulk part

$$
I_W(\beta)=\lim_{\beta\to0}{1\over{\rm Vol}\CG} \int d\eta [dx]^{D-1} \tr\left((-)^F\,e^{i\eta^A C^A}\,e^{-\beta H} \right).
$$
and a deficit part
$$
I_{deficit} = \int_0^\infty d\beta {d\over d\beta} \tr\left( (-)^F e^{-\beta
    H}\right)
$$
which can be rewritten as
$$
I_{deficit}= -\int_0^\infty d\beta \tr\left((-)^F H e^{-\beta H}
\right)=\int_{E>0} \{ \rho_+(E)-\rho_-(E)\}\ .
$$
From this expression it is obvious that the non-zero value of the deficit term
is given by the continuous part of the spectrum, as the discrete parts cancels
by supersymmetry. The total index can be rewritten as 
$$
I_W=I_W(0)+I_{deficit}\ .
$$
Following the above references, we rescale the field by
$$
\eta \to \beta\ X^0,\quad X^m \to  \ X^m,\quad
\Psi \to \Psi.
$$ 
In the $\beta\to 0$ limit it is possible to expand the trace over the bosonic
coordinates using the heat kernel expansion \refs{\rfYi,\rfSethiStern}, and
the bulk term can easily be rewritten as $I_W(0) = \lim_{\beta\to0} \CZ_{[D\to
  0]}^{(N)}$ with

\eqn\eRegulated{
\CZ^{(N)}_{[D\to0]} = {1\over\beta^{(N^2-1)(D-2+\CN/2)}} \int [dX]
[d\Psi] e^{-\beta \CS_{[D\to0]}}
}
where $\CS_{[D\to0]}$ is the zero-dimensional Matrix model action 

\eqn\eZeroeff{
\CS_{[D\to0]}={\beta\over 4\pi^2 g_sl_s^5} \Tr\left\{{1\over
    4}\left[X^\mu ,X^\nu\right]^2 +{1\over 2} \Psi^T\Gamma^0\Gamma^\mu [X^\mu ,\Psi] \right\},
}
obtained by dimensional reduction from the $D$ dimensional one. The upshot of
the expression~\eRegulated\ is that the overall power of $\beta$ cancels due
to identity of the bosonic transverse and the fermionic degrees of freedom, so
the $\beta\to0$ limit is well defined. In the previous path integral the
integration over the fermion has to be done with {\sl periodic\/} boundary
condition as enforced by the insertion of the $(-)^F$ operator in the
definition of the index.\foot{Hereafter, we call this contribution a partition
  function, but one should keep in mind that the fermions satisfy periodic
  boundary conditions.} In the final expression, we see that we have reduced
the model to a zero dimensional matrix model with an effective coupling
constant given by $g^2_{eff}=4\pi^2 g_sl_s^5/\beta$. 

Of course, in a rigorous computation along the lines of
\refs{\rfYi,\rfSethiStern} or \refs{\rfMooreNekrasovShatashvili},
all the elements of the matrices belonging to the non-Cartan part of the group
have to be included, but as we will see in the following, putting this
model on a two-torus and sending $\beta$ to zero {\sl before\/} computing the
integrals gives the correct answer. 
We want to use the non-trivial structure of the orbifold limit of
two-dimensional gauge field theory \refs{\rfDVV}. 
When taking this limit, the scaling of the fields is crucial, as the important
contributions have to come from fixed points of this orbifold space
\refs{\rfSeibergOrbifold}.  In our case the relevant scaling can be understood
as follows. 

We start from the zero-dimensional model~\eZeroeff\
and we compactify the coordinates $X^0$ and $X^9$ on a two-torus. Rewriting
those fields in a Fourier transform basis \refs{\rfTaylorCompact}\  $FT(X^{0,D-1})=2\pi l_s^2 D_{0,D-1}$
gives 
$$
\CS_B = {1\over 4g^2_{eff}} \int d^2\sigma\Tr\left[(2\pi l_s^2)^4 F^2+(2\pi l_s^2)^2 (DX^I)^2+
  [X^I,X^J]^2 \right].
$$
As the limit $\beta\to0$ is the same as $g^2_{eff}\to\infty$ in order to
exhibit the non-trivial dynamics of this limit, we rescale the fields as
$$
\CS_B^{rescaled}=\int d^2\sigma {(2\pi l_s^2)^4\over
  4g^2_{eff}}\Tr F^2+{1\over 4}\Tr(DX^I)^2+{g^2_{eff}\over4(2\pi l_s^2)^2}\Tr
[X^I,X^J]^2\ .
$$

We can now consider the infrared limit ($g^2_{eff}\to\infty$). In this
limit the gauge field decouples and we have to set the bosonic and
fermionic potentials to zero. The existence of flat directions gives rise to a moduli
space with an orbifold structure $(\RR^{D-2})^N/S_N$ where the group of
permutations $S_N$ is the Weyl group of $U(N)$.   
This is  a convenient way to
separate the dynamics of these matrix models, which split naturally into the
dynamics of two of the coordinates, the gauge field and an $X^9$ coordinate
for the $0+1$ Quantum mechanical models analysed in
\refs{\rfdeWitLuscherNicolai} or for the topological setting of
\refs{\rfMooreNekrasovShatashvili}.

Moreover, due to the supersymmetry of the model and the fact that we will
subtract the zero-modes of the fields (see the definition of the measure
of~(5.1)), the computation will {\sl not\/} suffer from any infrared
divergences as the regulator cancels automatically between the $D-2$ bosonic
coordinates and the $\CN/2$ fermionic ones.\foot{With the normalisations of
  reference~\refs{\rfKostovVanhove}, the contributions of the bosonic and
  fermionic Higgs fields cancel; this is, in particular, independent of the
  gauge coupling.} This is intimately linked to the underlying gauge symmetry
of the problem. A $U(N)$ matrix model {\sl without\/} the gauge symmetry has no
reason to be free of infrared divergences. In the end, we will be left with a
two-dimensional gauge field theory, as we will see in equation~(5.5).\foot{The
  finiteness of the matrix integral~\eRegulated\ for the $U(2)$ case was shown
  by the explicit computations of
  \refs{\rfKrauthNicolaiStaudacher,\rfGreenGutperleTwo,\rfYi,\rfSethiStern}.} 

\newsec{The Quasi-classical Evaluation of the Partition Function}
\seclab\quasiclassical

In order to perform the computation we put the previous Matrix Model on a
two-dimensional torus, and adopt the language of the reduction to two
dimensions of the ten-dimensional super Yang-Mills theory. We will follow
closely the notation and the logic of \refs{\rfKostovVanhove}, except that
the Yang--Mills coupling constant will be reintroduced. We will consider the
generalised setting of the model deduced by dimensional reduction from the
$D=3,4,6$ and 10 $U(N)$ super Yang-Mills theories. A subscript $[D\to d]$ will
indicated the dimension $D$ of the mother theory and the dimension $d$ of the
model under consideration. The computation will be done after having projected
all the fields onto the Higgs branch, described by the symmetric orbifold space
$(\RR^{D-2})^N/\CS_N$, and the zero-size limit of the torus will be taken to
get back to the zero-dimensional model of equation~\eDefIKKT. 

The partition function under study is
\eqn\eDefZ{
Z_{[D\to 2]}^{(N)}(\CT,g) = \int [\CD A][\CD X][\CD\Psi] \, e^{-\CS_{[D\to 2]}[A,X,\Psi]}.
}
The action is that of the Matrix String on a two-dimensional Minkowskian
torus~\eDefDVV
\eqn\esYM{
\CS_{[D\to 2]}={1\over \hat g_{YM}^2} \int d^2z \Tr\left[F_{ab}^2+\left(D_a
    X^I\right)^2 +i \bar\Psi \Gamma^a D_a\Psi - [X^I,X^J]^2 +
    \bar\Psi \Gamma^I [X^I,\Psi] \right],
}
with the coupling constant
$\hat g_{YM}^2 =g^2_{eff} RT=4\pi^2 g_sl_s^5 RT/\beta$. T-duality relates the
radii to the string coupling $g_s$ and the T-dual dilaton $\phi'$ constant by
$RTg_s=RT\exp(\phi)=\exp(\phi')l_s^2$; henceforth the gauge coupling is
expressed as $\hat g_{YM}^2=4\pi^2l_s^7\exp(\phi')/\beta$. Due to the
invariance of the theory under area-preserving diffeomorphisms of the
two-dimensional model, every quantity computed in this theory is a
function only of the parameter $RT\hat g^2_{YM}=4\pi^2
l_s^9 \exp(2\phi'-\phi)/\beta=4\pi^2
l_s^9/(g_s\beta)$. In the second equality we used the invariance under
area-preserving diffeomorphisms to set to zero the value of the T-dual
dilaton $\phi'$.\foot{This choice of unit corresponds to the limit $RT\to 0$
  and $l_s\to 0$ with $l_s^2/(RT)$ kept fixed.} This will be important when we
make contact with the supergravity theory in section~6.2.
The integration measure has the zero-modes deleted~\refs{\rfKostovVanhove}:
$$
\Psi^{(0)}_\alpha = {\Tr\int_{\CT^2} d^2z \Psi_\alpha\over \sqrt{NRT}},
\quad X^{(0)}_I = {\Tr\int_{\CT^2} d^2z X_I\over
  \sqrt{NRT}},\quad
\vartheta(\tau)= \Tr\int_0^R\! d\sigma A_\sigma(\sigma,\tau),\quad
A^{(0)}_\sigma = {\int_0^{mT}  d\tau \vartheta(\tau)\over \sqrt{NRT}}
$$
$$
[\CD A]=\CD A \delta\left(A^{(0)}_\sigma \over\sqrt{2\pi \hat g_{YM}^2}\right),\quad [\CD X ]=
\CD X  \prod_I \delta\left(X_I^{(0)}\over \sqrt{2\pi \hat g_{YM}^2}\right) ,\quad
[\CD \Psi]= \CD \Psi \prod_\alpha \, {\Psi_\alpha^{(0)}\over\sqrt{\hat g_{YM}^2}}.
$$
The world-sheet coordinates are $z= (\sigma,\tau)\in[0,R]\times [0,T]$. 


\subsec{The $\hat g^2_{YM}\to\infty$ limit}

The limit $\hat g^2_{YM}\to\infty$ (corresponding to $\beta\to0$) for the bulk
term of the Witten index projects the computation on the Higgs branch of the
model. In~\refs{\rfKostovVanhove} the field configuration in the infrared
limit of the model was worked out. In this limit  all matrices  $\cyrf = \{
D_a , X_I, \Psi_\alpha\}$ are simultaneously diagonalisable with a unitary
matrix $V(\sigma,\tau)$ such that 

$$\cyrf(\sigma, \tau)  =  V^{-1}(\sigma, \tau ) \cyrf^D(\sigma, \tau )V(\sigma, \tau )$$
where $\cyrf^D = {\rm diag}\{\cyrf_1,...,\cyrf_N\}$.
We therefore have
$$
\cyrf^D (R, \tau )= \hat T ^{-1} \cyrf^D ( 0, \tau )\hat  T ,\quad
   \cyrf^D (T , \sigma )= \hat  S^{-1}   \cyrf^D (0, \sigma)\hat   S  
$$
where $\hat S= V^{-1}(R, \tau ) V(0, \tau)$ and $\hat  T= V^{-1} (\sigma, T )
V(\sigma, 0)$.
By construction, $\hat S\hat T = \hat T \hat S $, and the matrices $\hat S$ and $\hat T$ 
represent two commuting permutations, $\hat s$ and $\hat t$, of the diagonal
elements. The permutations are given explicitly
as commuting elements $\hat s: i\to s_i $ and $ \hat t: i \to t_i $
 of the symmetric group $S_N$
$$\left[ \hat T ^{-1} \cyrf \hat T\right]_i = \cyrf _{t_i},\quad
\left[ \hat  S ^{-1} \cyrf \hat  S \right]_i = \cyrf _{s_i}
$$
The saturation of the fermionic zero modes restricts the computation to
a one-component covering of the torus, and the permutations are given by
\eqn\eBCj{
\hat s=\{ i\to i + m ({\rm mod} N)\} ,  \ \ \hat t=
\cases{ \{i \to i+ 1 ({\rm mod} m)\} & if $j=0$\cr
 \{i \to i- j ({\rm mod} N)\} & if $j=1, ..., n-1$\cr}
}
The topological sectors of the partition function are classified by the
permutations
$$
  \int[\CD A ][\CD X] [\CD \Psi] \rightarrow {1\over N!} 
 \sum_{\hat s \hat t = \hat t \hat s}
\int  [\CD A^D][\CD X^D] [\CD \Psi^D]\ ,
$$
and the partition function after integration over the bosonic Higgs fields and
the fermions can be written as
\eqn\eZdvv{
\CZ_{[D\to 2]}^{(N)} (RT\hat g^2_{YM}) = { (N-1)!\over N!} \sum_{mn=N}\sum
_{j=0,\dots,n-1}\,\delta^{\rm susy}_{m,n} \CZ_{[m, n; j]}
=  \sum_{m|N}{1\over m}\,\delta^{\rm susy}_{m,n}\CZ_{mn}\ .
}
The function $\delta^{\rm susy}_{m,n}$ is a function of the boundary conditions
induced by the integration over the fermions, which will be discussed in the
next subsection. It was shown in~\refs{\rfKostovVanhove} that $\CZ_{mn}$
reduces to the partition function of a $U(1)$ gauge theory defined on the torus
of area $NRT$ with periods $( mT ,jR)$ and $(0,  nR)$ because, having
subtracted the zero modes, supersymmetry ensures that the bosonic and fermionic
determinants cancel exactly, with the result\foot{The fact that the $U(N)$
  partition function can be decomposed as a sum over $U(1)$ partition functions
  with respect to $S_N$-cycle decomposition over a torus of size extended by
  the length of the cycle, was known by I.K.~Kostov and the author and appeared
  in \refs{\rfVanhovePhd}; this was independently discovered by
  \refs{\rfBillo}.}  
\eqn\eReduction{
\CZ_{[D\to 2]}^{(N)}(RT\hat g^2_{YM})=  {1\over T} \sum_{m|N}{1\over 
m^2}\ \delta^{\rm susy}_{m,n}\sum_{E\in\ZZ} \exp\left(-{E^2\over2 \hat g^2_{YM}
  RTN}\right).
}
Decompactifying with the limit $RT\to0$ and  setting $\SV\Definition
(RT)^{{7\over 2}N^2}\ {\rm Vol}(SU(N)/\ZZ_N)$ to be the overall volume factor,
the partition function is
\eqn\eMps{
\CZ_{[D\to 0]}^{(N)}=  \SV\times \left(RT\hat g^2_{YM}\right)^{7/2}
\, \sum_{m|N}{1\over m^2}\ \delta^{\rm susy}_{m,n}. 
}
The factor $RT\hat g^2_{YM}$ comes from the normalisations of the $U(1)$ part of the matrices.

\subsec{Constraints from Supersymmetry}
\subseclab\constraints

 We have to compute the integral over the fermionic variables
\eqn\eZfermions{
\left.Z[A]\right|_{mn} = \int_{b.c.} [\CD\Psi^D] \ \exp\left(\bar\Psi^D
  i\not\!\!D\Psi^D \right)
}
with boundary conditions~\eBCj. To compute the integral over the fermions it
is crucial to define the measure. The measure $[\CD\Psi^D]$ is defined using
the mode expansion of the $\Psi^i$ with respect to the kinetic operator
\refs{\rfFujikawa}. We consider 
$$
\eqalign{
i\not\!\! D \phi_n &= \lambda_n \phi_n\cr
\Psi(\sigma,\tau) = \sum_n \phi_n(\sigma,\tau) a_n&,\quad
\bar\Psi(\sigma,\tau) = \sum_n \phi_n^\dagger(\sigma,\tau) b_n\cr
}
$$
Hence
$$
[\CD\Psi^D] \Definition \prod_n \, da_n db_n
$$
In the case  with $\CN=16$ real supercharges the kinetic operator splits as
$\Sl^t D \Sl+ \Sr^t \bar D \Sr$, with respect to the two fermionic Spin(9)
representation of a  Majorana-Weyl spinor
$\Psi_{16} = \left(\Sl\ \Sr\right)$. 
The index $\alpha$ runs over the $\huits$ representation, and $\da$
over the $\huitc$ representation of Spin(9); see \refs{\rfBaake} and Appendix~B for
representations of the Clifford algebra.
So the mode expansion for the light-cone fermions $S$ is $S = \sum_n
\phi_n(\sigma,\tau) a_n$ and all boundary conditions are satisfied.

For the theories with $\CN=4,8$ real supercharges, such a chiral factorisation
is not allowed as the fermions only satisfy the Majorana condition. The
kinetic operator for the case with $\CN=4$ real supercharges reads

$$
{}^t\chil\alpha
\left(i\sigma^3\partial_1\chil\alpha+\sigma^2\partial_0\chir\alpha\right)
+{}^t\chir\alpha\left(i\sigma^3\partial_1\chil\alpha+\sigma^2\partial_0\chir\alpha\right)
$$
In the Higgs phase $\chil\alpha={\rm diag}(\chil\alpha^l)$ ($l=1,\dots,N$) the
equations of motions read for each eigenvalue

$$
\eqalign{
\partial_1\chil1-\partial_0\chir2 =0,&\quad-\partial_1\chil2+\partial_0\chir1=0\cr
\partial_1\chir1-\partial_0\chil2 =0,&\quad-\partial_1\chir2+\partial_0\chil1=0\cr
}
$$
These equations imply that all fermions satisfy the Klein-Gordon equation
$(\partial_0^2-\partial_1^2)\chi=0$, and due to the reality condition
$\chil\alpha=(\chil\alpha)^*$ and $\chir\alpha=(\chir\alpha)^*$ their mode
expansion is  

$$
\chil\alpha = \sum_n
e^{-n\tau}\left(a_{n,\alpha}\,e^{-in\sigma}+a_{n,\alpha}^*
  e^{in\sigma}\right).
$$
Therefore, only trivial boundary conditions in the $\sigma$ direction are
possible. This means that only the configuration with $n=1$ ({\sl i.e.} $S=1$)
and $m=N$ contribute to the partition function~\eMps.  

From this analysis, the constraints from the integration over the fermions
are summarised by 

\eqn\eDelta{
\delta^{\rm susy}_{m,n}=\cases{n=1,\, m=N& when $\CN=4,8$ \cr mn=N &
  when $\CN=16$,}
}
giving the final result
\eqn\eZfinal{
\CZ_{[D\to0]}^{(N)} = \SV\times \left(RT\hat g^2_{YM}\right)^{7/2}\times \cases{1/N^2 & for
  $\CN=4,8$ {\sl i.e.} $D=4,6$ \cr\sum_{m|N}1/m^2 & for $\CN=16$ {\sl i.e.} $D=10$.}
}

The special case with two real super-charges $(D=3)$ is treated independently
in the Appendix~A. 

\newsec{Contact with threshold corrections}
\subsec{The Heterotic/Type~I threshold corrections}

There is another example where the Matrix string setting is helpful to analyse
the configuration of fermionic zero modes, namely the case of the $\CF^4$ and
$\CR^4$ terms in the effective action for the $Spin(32)/\ZZ_2$ type~I
theory. The contribution of the complete effective action for the type~I string
on $\RR^{1,7}\times \CT^2$ consists of perturbative and non-perturbative terms
induced by the wrapped Euclidean\foot{This wrapped D1-brane should be
  Euclidean, so a Wick rotation on the world-sheet of the previous
  two-dimensional model is necessary. The Majorana-Weyl fermions are now
  converted to complex Weyl fermions. See \refs{\rfDhokerPhongFermions}, for
  instance, for an explanation of how to handle this case.} D1-brane on a
two-torus included in the $d$-torus.\foot{While this paper was being proofread
  the preprint \refs{\rfGutperleIprime} appeared with some related comments.}
These contributions can be completely evaluated because they are all mapped
together to the one-loop amplitudes of the heterotic string on $\RR^{1,7}\times 
\CT^2$. The non-perturbative part on the type~I side can be written in the
compact form as  \refs{\rfBFKOV,\rfKiritsisObers}

$$
\CI^{inst} = - {V^{(10)}\over 2^8 \pi^4}
\sum_{K=1}^\infty \, {e^{2i\pi K T}\over T_2}\ \CH_{K}[\CO\hat \CA](U)+ c.c.
$$
where $\CO=1+\cdots$ is a differential operator, whose action is induced by the
non-holomorphic terms in the elliptic genus $\hat \CA$ and gives rise to a
finite number of higher loop effects around the D-instanton.  Here
$$
\CH_K[\hat \CA](U)= {1\over K} \sum_{mn=K\atop 0\leq j < n} \hat\CA\left(j+
  m U\over n \right)
$$
is the Hecke operator of rank $K$ acting on the modular elliptic genus $\hat
\CA$, whose gauge field part  
$$
\hat\CA(U)= t_8 \tr F^4 + {1\over 2^9 3^2}\left[ {E_4^3\over \eta^{24}}+ {\hat
      E_2^2 E_4^2\over \eta^{24}}
-2{\hat E_2 E_4 E_6\over \eta^{24}} -2^7 3^2\right] t_8 (\tr F^2)^2
$$
only will be needed for this discussion. The coefficient of the $\tr F^4$ terms is explicitly given by 
$$
\CH_K(1) = \sum_{m|K} {1\over m}.
$$ 
In this case the pertinent (T-dual) Matrix Model is the $(8,0)$ quantum
Mechanical model considered in
\refs{\rfBanksSeibergSilverstein,\rfDanielssonFerretti,\rfBachasGreenSchwimmer}

$$
\eqalign{
\CS_{[10\to1]}={1\over2 g}&\Tr\left({1\over 2} (DX^m)^2 - {1\over 2}
  (D\Phi)^2+{1\over 4} \left[X^m,X^n\right]^2+{i\over2} \Theta D\Theta -
  {i\over2} \Theta[\Phi,\Theta] - {1\over 2}[\Phi,X^m]^2\right.\cr
&\left.- {i\over2}\lambda D\lambda -
  {i\over2}\lambda[\Phi,\lambda] + i
 X^m\gamma^m_{a\dot a}\{\Theta_a,\lambda_{\dot a}\} + i \chi^I D\chi^I + i\chi^I\Phi\chi^I +i m^{IJ} \chi_I\chi_J
\right)
}
$$
The coordinates $X$ and $\Theta$ are in the symmetric representation of
$SO(N)$, while $\Phi$, $\lambda$ and the gauge connection $A_0$ are in the
antisymmetric representation. $\chi^I$ $(I=1,\dots,2\SN)$ are in the real
representation $(K,2\SN)$ of $SO(K)\times  SO(2\SN)$. We now put the coordinate
$\Phi$ on a circle of radius $l_s^2/R$, and convert it into a second gauge
connection component. We obtain a two-dimensional gauge model. This model
presents several important differences relative to the one studied
previously. Only the matrices $X$ and $\Theta$ have Abelian degrees
of freedom. The gauge coordinates do not have any Abelian quantum numbers
because the group is $SO(K)$. Then the measure for the gauge connection is
simply   
$$
\left[ \CD A\right] = \CD A.
$$
We {\sl assume\/}\foot{The saturation of the fermionic zero-modes does not
  appears as easily as for the model~\eDefZ. And the infrared finiteness of
  the model is not obvious. Equivalently, the $\beta\to0$ limit of the bulk
  term for the Witten index of this type~I$^\prime$ model is not as trivial as
  before. But the existence of a sound limit is linked with the supersymmetry
  content of the model, and restricts the integral to be computed. A more
  rigorous derivation is a little more subtle and is deferred to a future
  publication \refs{\rfVanhoveWork}.} that we can project the theory onto the
classical moduli space,  $\CM =
\left(\RR^{1,7}\times \CT^{2}\right)^K/(S_{K/2}\times \ZZ_2^{K/2})$
\refs{\rfRey,\rfLowe,\rfHorava}, and decouple the gauge degree of freedom from
the Higgs field dynamics. Since we are only interesting in the measure factor
$\mu (K)$, the gauge field degrees of freedom are the only ones needed. A
computation similar to the one done in section~\quasiclassical\ gives    

$$
\mu _{type~I} (K) = \CH_K(1).
$$

\subsec{The D-instantons}
\subseclab\secconnec

In order to make  contact with D-instanton corrections of the action~\eAction\
of section~\secsugra, it is necessary to specify the normalisations of the
Matrix Model since for the D-instanton there are different from those of the
gauge theory model of section~\quasiclassical. We identify the $U(1)$ part of
the matrix coordinates with the superspace coordinates of section~\secsugra,  
$$
x_\mu = \Tr\left( A_\mu \over N\right), \quad \theta=\Tr\left(\Psi \over N\right).
$$
Moreover it is necessary to eliminate the volume of the gauge group, $\SV$;
then the partition function of the D-instanton Matrix Model is defined by  
\eqn\eZdinst{
\SV\, Z_{\rm D-ins}^{(N)} \Definition \int dX_0\prod_{\mu =1}^9 \, dX_\mu  \,
\delta\left(\Tr X_\mu  \over N\right) \ \prod_{\alpha=1}^{16} \, d\Psi_\alpha \,
\delta\left(\Tr \Psi_\alpha\over N\right) \  \exp(-S_{\rm D-ins}), 
}
$$
S_{\rm D-ins}= {1\over g_{ins}} \Tr\left({1\over 4}[X_\mu ,X_\nu]^2 +{1\over
    2}\Psi^T \Gamma_\mu [X^\mu ,\Psi]\right)
$$
where $g_{ins}=g_0=g_s/l_s^4$. Using $\hat g^2_{YM}=g_0RT$ and $RT\hat
g^2_{YM}=g_s$ in the formulae of section~\quasiclassical, we get
\eqn\eZdins{
Z^{(N)}_{\rm D-ins}= \mu (N) \left(N e^{-\phi}\right)^{-7/2}.
}
It should be remarked that this formula is independent of the string scale
$l_s$. Depending on which component of the superfield~\ePhi\ this D-instanton
background couples to, a different power of $N\exp(-\phi)$ appears, namely
$w+1/2$, according to~\eModularexp. This can be derived by explaining how to
couple the supergravity field to the previous Matrix model. We have seen that
the computation of the measure factor $\mu (N)$ can be reduced to the
contribution from the $U(1)$ part of the two-dimensional gauge field (see
equation~\eReduction), a feature particularly difficult to analyse in the
zero-dimensional version of the model. The various supergravity states which
appear in the decomposition of the superfield~\ePhi\ are now seen as states of
the Abelian part of the Matrix Model, and it is possible to construct them as
eigenstates of the centre of mass Hamiltonian \refs{\rfPlefkaWaldron} by
decomposing the representations of $SO(9)$ under $SO(7)\times U(1)$.\foot{see
  \refs{\rfdeWitLouis} for a clear lecture on the subject.}

It is natural to represent the external supergravity states by Wilson point observables \refs{\rfThompson}
$$
W_{Wilson\ point}\Definition \Tr_{adjoint}\exp\left({1\over \hat g^2_{YM}}\int_\Sigma
  d^2z \star F \right).
$$
This operator is the only observable which decompactifies correctly in the
limit $R,T\to 0$. We promote the $U(1)$ part of this curvature to a superfield
$\FF^{U(1)}(x^\mu ,\theta^\alpha)$ for the Abelian supersymmetry
transformation~\eAbeliansuzy\ and identify the  superfield $\Phi$ of~\ePhi\
with $l_s^4\star\FF^{U(1)}=\Phi$. We assign $U(1)_B$ weight $+1/2$ to
the supercharges associated with the supersymmetry
transformations~\eAbeliansuzy\ and  weight $-1/2$ to the supercharges
associated with the transformations $\delta^1_\epsilon$ of the $SU(N)$ part of
the coordinates. Units are as before: the supergravity coordinates $X$ have
dimension $l_s^2$ times the gauge field connection ones. The coupling of the
Matrix states with the supergravity external states is given by insertions of

\eqn\eWilson{
W_{Wilson\ point}= N\times \exp\left({1\over g_s}\Phi+c.c.\right),
}
The term with $U(1)_B$-weight $w$ is now given by a correlation of $w+4$
insertions of the operator $W_{Wilson\ point}$ where we pick the
sixteen-$\theta$s term. The gauge symmetry fixes the power of the string
coupling constant in~\eWilson\ in such a way that the sixteen-$\theta$s term is
proportional to $(N/g_s)^{w+4}$, which multiplies~\eZdins. This is exactly what
is needed to reproduce the result of~\eModularexp.

\newsec{Discussion}

\subsec{The large $N_c$ limit}

The conjecture about the equivalence between the supergravity theory and the
super-conformal Yang-Mills theory on the boundary space relies heavily on the
large-$N_c$ limit on the Yang-Mills side. Here we have only discussed
finite-$N_c$ computations but we can see from these that a large-$N_c$ limit
can be defined. The guideline for this is that we have to keep a {\sl fixed
  number of fermionic zero-modes\/} irrespective of the rank of the gauge
group. If we consider a configuration of $K$ long strings, restricted to join
forming long strings of size a multiple of $N_c$, but {\sl not to split\/}, all
the dynamics is embedded in a $U(N_cK)$ group. From the previous analysis, of
the zero-dimensional model for the D-instantons and the computation of
section~\secconnec,  we deduce that the measure factor for the interactions is
$\mu (K)$, and the factor of $N_c$ appears as an overall power. It is now safe to
take the large-$N_c$ limit keeping $K$ finite. 

If we start from a configuration of fields decomposed as $A_{N_cK}\Definition
A_K\otimes \II_{N_c}$ for a $U(N_cK)$ model, the action $S_{\rm D-ins}$ gets an
overall $N_c$ factor, so for such a field configuration

\eqn\eScaling{
\eqalign{
\left.S_{\rm D-ins}\right|_{N_cK}&= N_c \left.S_{\rm D-ins}\right|_K\cr
&={N_cl_s^4\over g_s} \Tr_{K\times K}\left({1\over 4}[X_\mu ,X_\nu]^2 +{1\over
    2}\Psi^T \Gamma_\mu [X^\mu ,\Psi]\right)
}}
and since we do not get any extra factor of $N_c$ from the measure,  it follows
that
\eqn\eLimit{
Z^{(N_cK)}_{\rm D-ins}=N_c^{-7/2} Z^{(K)}_{\rm D-ins}=N_c^{-7/2}\mu (K)
\left(Ke^{-\phi}\right)^{-7/2}\ .
}
The Wilson point observables are now given by
$$
W_{Wilson\ point}= K\times \exp\left({1\over g_s}\Phi+c.c.\right)\ ;
$$
because the interaction occurs only between long-strings of 
length a multiple of $N_c$, we do not get any extra power of $N_c$ by inserting them:
\eqn\eCorrelation{
\left\langle (W_{Wilson\ point})^{w+4}\right\rangle =
N_c^{1/2}\times \left(1/N_c\right)^4\times \mu (K)\left(Ke^{-\phi}\right)^{-7/2+w+4}.
}
This means that we can take a large-$N_c$ limit, with the instanton number $K$
fixed, in a well-defined way. This splitting of the fields means that we
restrict the integration in the matrix model to occur only between long
strings with length {\sl at least $N_c$}. This can be made much more
rigorous by embedding a  $U(K)$
instanton constructed by Giddings, Verlinde and Hacquebord
\refs{\rfGiddingsVerlindeHacquebord,\rfBonora}, tensored with the diagonal
matrices doing cycles of length $N_c$ along the lines of
\refs{\rfWynter}.\foot{This configuration smoothes the gauge interactions out of
  the infrared limit at finite $\hat g^2_{YM}$.}
Finally, the scaling~\eScaling\ shows that the combination $N_c\alpha'{}^2$
appears naturally, as a reminiscence of the scaling $\alpha'{}^{-1}\propto
N_c^{1/2}$ of the AdS/CFT correspondence \refs{\rfAdSCFT}.
The overall power of $\alpha'$ in~\eLimit\ is not correct, due to our inability
to derive the relative normalisation with respect to the
kinetic Einstein-Hilbert term of the supergravity theory.

\subsec{D-instantons Loop expansion}

That we were not able to get the full expansion of the modular
form~\eModularexp\ and~\eCoeff\ from our Matrix model analysis is certainly
due to the fact that we are only getting the physics of the linearised version
of the supergravity theory, which gives only the dominant term in the
instantons expansion. This is not surprising since the $Sl(2,\ZZ)$
  symmetry of the type~IIb theory is not explicit in the model. It may be
  possible to understand how this symmetry can appear along the lines of
  \refs{\rfIntriligator} by considering the constraints from the $U(1)_B$
  weight of the fields.
Moreover it would certainly be worth analysing how much information we can get
about these $\alpha'{}^3$ corrections by deforming the superspace analysis. We
hope to return to this problem in a future publication.   


\bigbreak\bigskip\bigskip\centerline{{\bf Acknowledgements}}\nobreak

It is a pleasure to thank Michael Green for many valuable discussions and
interest in that work, and Ivan Kostov for enjoyable collaboration at the
beginning of this work. The author warmly thanks Tim Hollowood and Michael
Mattis for useful comments on a misconception in the previous version of the
manuscript. We would like to acknowledge the LPTHE of Jussieu for hospitality
where part of these reflections were done, and the organiser of the {\sl 65{\`e}me rencontre de Strasbourg\/} for an invitation to present them. 
The author would like to dedicate this article to the bakery {\sl Fitzbillies\/}
for providing his daily chocolate fudge cake before it was destroyed in a fire.
This work was supported by a PPARC research fellowship.


\vfill\supereject

\appendix{A}{The $\CN=2$ real supercharges case}

For the case of $\CN=2$ real supercharges the two-dimensional model~\esYM\ is
composed of gauge field $A_{\sigma,\tau}$ and one bosonic Higgs $X$, and
its real bispinor partner. The only potential is the one induced by the
covariant derivative so it will not be necessary in that case to take the
infrared limit of the model. The measures are defined as in
section~\quasiclassical. 

The field $X(\sigma,\tau)$ is conjugate to an element in the Cartan
subalgebra of $U(N)$
$$
X(\sigma,\tau)=V^{-1}(\sigma,\tau) X^D(\sigma,\tau) V(\sigma,\tau)\ ;
$$
likewise for the fermion due to space-time supersymmetry
$$
\Psi(\sigma,\tau)=V^{-1}(\sigma,\tau) \Psi^D(\sigma,\tau) V(\sigma,\tau).
$$
Doing the gauge transform

$$
A_\alpha(\sigma,\tau)\to V^{-1}(\sigma,\tau)
\left(\hat A_\alpha(\sigma,\tau)+\partial_\alpha \right)V(\sigma,\tau)\ ,
$$
we can now integrate the diagonalized Higgs fields.  Once again, the
determinants (without the zero modes) cancel due to supersymmetry, leaving just
a constraint on the gauge field configurations from the equations of motion of
the fermions
$$
Z_{[3\to 2]}^{(N)}= \int [\CD \hat A]\, \delta^{\rm susy}_{[\hat A]} \ 
e^{-{1\over 4\hat g^2_{YM}}\int F^2}.
$$
The equations of motion for the fermions reads
\eqn\eFermvt{
\not\!\partial \Psi^D + [ \not\!\!\hat A,\Psi^D]=0.
}
Because the fermions are in the Cartan torus the group indices of the
connection in $[\hat A_\alpha,\Psi^D]$ can be restricted to the orthogonal
(w.r.t. the Cartan metric) complementary, $n$, of the Cartan subalgebra of
$u(N)$, $[\hat A_\alpha^n,\Psi^D]$. Moreover $[A_\alpha^n,\Psi^D]$ belongs to
$n$, so from~\eFermvt\ we deduce that  
$$
\not\!\partial \Psi^D=0, \quad [ \not\!\!\hat A^n,\Psi^D]=0.
$$
The second equation implies that $\hat A^n_\alpha =0$.  Henceforth, the
configurations of the fields are classified as before (see equation~\eZdvv\ and
section~\constraints) with the result   
\eqn\eZtroistodeux{
\CZ_{[3\to 0]}^{(N)}=\SV\times \left(RT\hat g^2_{YM}\right)^{7/2} {1\over N^2}.
}

\vfill\supereject

\appendix{B}{Representation of the real Clifford Algebra}

We list the $d\times d$ irreducible representations of the real
Clifford algebras. The spinor are chosen real, $\psi=\psi^*$, the
charge conjugation matrix is $C=\Gamma^0$ and $(\Gamma^\mu 
)^*=\Gamma^\mu $.

$$
\{\Gamma^\mu ,\Gamma^\nu\}=2g^{\mu \nu} \quad (\mu 
,\nu)\in\{1,2,\dots,D\}^2,\quad sign(g)=(-1,+1,\dots,+1)
$$

The Pauli matrices are

$$
\sigma_1=\pmatrix{0&1\cr1&0}\quad \sigma_2=\pmatrix{0&-i\cr
  i&0}\quad\sigma_3=\pmatrix{1&0\cr0&-1}
$$
and we define $\epsilon=i\sigma_2$.

\subsec{$\CN=2$, $d=3$}
$$
\Gamma^0=\epsilon,\quad
\Gamma^1=\sigma_3,\quad
\Gamma^2=\sigma_1.
$$

\subsec{$\CN=4$, $d=4$}


The basis used by \refs{\rfBaake} is

$$
\Gamma^0=\pmatrix{0&\sigma_3\cr-\sigma_3&0}=\sigma_3\otimes\epsilon,\quad
\Gamma^1=-1\otimes\sigma_1,\quad
\Gamma^2=-\epsilon\otimes\epsilon,\quad \Gamma^3=1\otimes \sigma_3,
\quad\Gamma_5=-i\sigma_1\otimes\epsilon
$$


\subsec{$\CN=16$, $d=16$}

We choose a basis well adapted
for the $\huits\oplus\huitc$ decomposition of the representations of
Spin(9). All gamma matrices are sixteen-dimensional square matrices.
$$
\Gamma^9=1\otimes1\otimes1\otimes\sigma_3={\rm
  diag}(1^8,-1^8),\quad \Gamma^i=\pmatrix{0&\gamma^i_{a\dot a}\cr
  \gamma^i_{\dot b b}&0}\quad i=1,\dots,8
$$
In this basis there is no $\Gamma^0$ matrix. The $8\times 8$ $\gamma^i$ matrices are defined by

\eqn\egamma{
\eqalign{
\gamma^1=\epsilon\otimes1\otimes1,\quad\gamma^2=\sigma_3\otimes\epsilon\otimes\sigma_3&,\quad
\gamma^3=\sigma_1\otimes\sigma_3\otimes\epsilon,\quad\gamma^4=\sigma_3\otimes\epsilon\otimes\sigma_1\cr
\gamma^5=1\otimes\epsilon\otimes1,\quad\gamma^6=\sigma_3\otimes1\otimes\epsilon&,\quad
\gamma^7=1\otimes\sigma_1\otimes\epsilon,\quad\gamma^8=1\otimes1\otimes1\cr
}}
We have $\Gamma^i=\gamma^i\otimes\epsilon$ for $i=1,\dots,7$ and
$\Gamma^8=\gamma^8\otimes\sigma_1$. The full 32-dimensional gamma matrices are
obtained  by $\Gamma^\mu = \Gamma^\mu \otimes \sigma_1$.


\listrefs

\bye